\documentclass[11pt]{cernrep}
\usepackage{graphicx,axodraw}
  \renewenvironment{thebibliography}[1]{%
    \begin{oldthebibliography}{#1}%
      \setlength{\parskip}{0ex}%
      \setlength{\itemsep}{0ex}%
  }%
  {%
    \end{oldthebibliography}%
  }

\def\gsim{\mathrel{\raise.3ex\hbox{$>$\kern-.75em\lower1ex\hbox{$\sim$}}}}

\begin{document}
\begin{titlepage}
\def\thefootnote{\fnsymbol{footnote}}       % symbols for footnotes

\begin{center}
\mbox{ }

\end{center}
\vskip -3.0cm
\begin{flushright}
\Large
\vspace*{-2cm}
\mbox{\hspace{11.85cm} hep-ph/0312301}    \\
\mbox{\hspace{11.65cm} TSL/ISV-2003-0274} \\
\mbox{\hspace{11.85cm} SHEP-03-38}        \\
\mbox{\hspace{12.0cm} December 2003}
\end{flushright}
\begin{center}
\vskip 2.0cm
{\boldmath \Huge\bf
{The \boldmath$\rm p\bar p \rightarrow tbH^\pm$\unboldmath\ 
        Process 
\smallskip
at the Tevatron in HERWIG and PYTHIA Simulations}
}
\vskip 2cm
{\LARGE\bf Johan Alwall$^1$, 
           Catherine Biscarat$^2$, 
           Stefano Moretti$^3$, \\ 
           Johan Rathsman$^1$ and 
           Andr\'e Sopczak$^2$ \\
\bigskip
\bigskip
\bigskip
\Large
$^1$Uppsala University, Sweden;
$^2$Lancaster University, UK;   \\
$^3$Southampton University, UK
}

\vskip 2.5cm
\centerline{\Large \bf Abstract}
\end{center}

\vskip 2.2cm
\hspace*{-2cm}
\begin{picture}(0.001,0.001)(0,0)
\put(,0){
\begin{minipage}{17cm}
\Large
\renewcommand{\baselinestretch} {1.2}
Charged Higgs boson production in association with a top quark could be 
the first indication of the existence of Higgs particles. 
The Tevatron Run-II started data-taking in April 2001 at 
$\sqrt s=1960$~GeV and could probe the existence of a charged Higgs boson
beyond the current mass limit.
We study the $\rm p\bar p \rightarrow tbH^\pm$ production process
with Monte Carlo simulations in HERWIG and PYTHIA, comparing expected cross
sections and basic selection variables.
\renewcommand{\baselinestretch} {1.}

\normalsize
\vspace{3.5cm}
\begin{center}
{\sl \large
Contribution to the Les Houches workshop ``Physics at TeV Colliders'', 
26 May -- 6 June, 2003
\vspace{-3cm}
}
\end{center}
\end{minipage}
}
\end{picture}
\vfill

\end{titlepage}

%%%%%%%%%%%%%%%%%%%%%%%%%%%%%%%%%%%%%%%%%%%%%%%%%%%%%%%%%%%%%%%%%%%%%%%%%%%

\newpage
\thispagestyle{empty}
\mbox{ }
\newpage
\setcounter{page}{1}
\pagestyle{plain}
%%%%%%%%%%%%%%%%%%%%%%%%%%%%%%%% AS end %%%%%%%%%%%%%%%%%%%%%%%%%%%%%%%%%%%

 \title{THE \boldmath$\rm p\bar p \rightarrow tbH^\pm$\unboldmath\ 
        PROCESS AT THE TEVATRON IN HERWIG AND PYTHIA SIMULATIONS}
\author{Johan Alwall$^1$, Catherine Biscarat$^2$, Stefano Moretti$^3$, Johan Rathsman$^1$ and Andr\'e Sopczak$^2$\thanks{~~E-mail: Andre.Sopczak@cern.ch}}
\institute{$^1$Uppsala University, Sweden; $^2$Lancaster University, UK; $^3$Southampton University, UK}
\maketitle
\begin{abstract}
Charged Higgs boson production in association with a top quark could be 
the first indication of the existence of Higgs particles. 
The Tevatron Run-II started data-taking in April 2001 at 
$\sqrt s=1960$~GeV and could probe the existence of a charged Higgs boson
beyond the current mass limit.
We study the $\rm p\bar p \rightarrow tbH^\pm$ production process
with Monte Carlo simulations in HERWIG and PYTHIA, comparing expected cross
sections and basic selection variables.
\end{abstract}

\section{INTRODUCTION}
Charged Higgs bosons are predicted by non-standard models, for example Two-Higgs Doublet
Models such as the Minimal Supersymmetric 
Standard Model (MSSM).
Thus, their detection and the measurement of their properties 
(such as the mass which is not predicted by any model) play an important r\^ole in the 
investigation of an extended Higgs sector and in the understanding of the 
generation of particle masses.
The current limit on the charged Higgs boson mass is set by the LEP experiments
at 78.6~GeV, independent of the Higgs boson decay branching 
fractions~\cite{leplimit}.
At the Tevatron, charged Higgs bosons could be discovered for masses well 
beyond this limit.

If the charged Higgs boson mass $m_{\rm H^\pm}$ satisfies 
$m_{\rm H^\pm} < m_{\rm t} - m_{\rm b}$, where $ m_{\rm t}$ is the top quark mass and 
$ m_{\rm b}$ the bottom quark mass,
it could be produced in the decay of the top quark $\rm t \rightarrow bH^+$.
This so-called on-shell top approximation
($\rm q\bar q$, $\rm gg \rightarrow t\bar t$ with $\rm t \rightarrow bH^+$)
was previously used in the event generators. Throughout this paper this process is denoted by
$\rm p\bar p \rightarrow t\bar t \rightarrow tbH^\pm$.
Owing to the large top decay width ($\rm \Gamma_{\rm t} \simeq 1.5$~GeV) and because of 
the additional diagrams which do not proceed via direct $\rm t\bar t$ 
production~\cite{Borzumati:1999th,Miller:1999bm},
charged Higgs bosons could 
also be produced beyond the kinematic top decay threshold. 
The importance of these effects in the threshold region was emphasized in the 
previous Les Houches proceedings~\cite{Cavalli:2002vs} and the 
calculations~\cite{Borzumati:1999th,Miller:1999bm} are implemented in 
HERWIG\,\cite{Corcella:2000bw,Corcella:2002jc,Moretti:2002eu}\,and 
PYTHIA\,\cite{pythia}\footnote{HERWIG release version 6.505 and 
inclusion in a future official PYTHIA version.}. 
The full process is referred to as $\rm p\bar p \rightarrow tbH^\pm$.
Examples of the graphs contributing to the $\rm p\bar p \rightarrow \bar t bH^+$
process are~\cite{Guchait:2001pi}:

\vspace*{-1.4cm}
\begin{equation}
%Q1:
\begin{picture}(120,30)
\SetScale{0.8}
\SetWidth{0.6}
\SetOffset(2,-62.5)
\ArrowLine(30,65)(15,50)
\ArrowLine(15,80)(30,65)
%\Text(10,45)[]{$\rm \bar p$}
%\Text(10,85)[]{p}
\Gluon(30,65)(60,65){3}{3}
\ArrowLine(75,50)(60,65)
\ArrowLine(60,65)(75,80)
\DashLine(70,75)(85,75){2}
\Text(78,60)[]{\small$ \rm H^+$}
\Text(66,40)[]{\small $ \rm \bar t$}
\Text(66,70)[]{\small$ \rm b$}
\end{picture}
%G7:
\begin{picture}(120,40)
\SetScale{0.8}
\SetWidth{0.6}
\SetOffset(2,-62.5)
\Gluon(15,50)(30,65){3}{3}
\Gluon(30,65)(15,80){3}{3}
%\Text(10,45)[]{\small 2}
%\Text(10,85)[]{\small 1}
\Gluon(30,65)(60,65){3}{3}
\ArrowLine(75,50)(60,65)
\ArrowLine(60,65)(75,80)
\DashLine(70,75)(85,75){2}
\Text(78,60)[]{\small$ \rm H^+$}
\Text(66,40)[]{\small $ \rm \bar t$}
\Text(66,70)[]{\small$ \rm b$}
\end{picture}
\begin{picture}(120,30)
\SetScale{0.8}
\SetWidth{0.6}
\SetOffset(0,-60)
\Gluon(45,75)(30,90){3}{3}
\Gluon(30,40)(45,55){3}{3}
%\Text(25,95)[]{\small 2}
%\Text(25,35)[]{\small 1}
\ArrowLine(60,40)(45,55)
\ArrowLine(45,55)(45,75)
\ArrowLine(45,75)(60,90)
\DashLine(45,65)(60,65){2}
\Text(58,55)[]{\small$ \rm H^+$}
\Text(53,35)[]{\small $ \rm \bar t$}
\Text(53,75)[]{\small$ \rm b$}
\end{picture}
\vspace*{1.2cm}
\end{equation}
The t-channel 
graph is one example of a diagram which does not proceed via 
$\rm t\bar t$ production.
This graph contributes to enhanced particle production in the forward 
detector region.

A charged Higgs boson with $m_{\rm H^\pm} < m_{\rm t}$
decays predominantly into a $\tau$ lepton and
a neutrino. 
For large values of $\tan \beta$ ($\gsim$ 5), the ratio of the vacuum
expectation values of the two Higgs doublets, this branching ratio
is about 100\%.
The associated top quark decays predominantly into a W boson or a second charged Higgs boson,
and a b-quark. 
The reaction
\begin{equation}
\rm p\bar p \to  tbH^\pm~~~(t\to bW^\mp)~~~(H^\pm \to \tau ^\pm \nu_{\tau})
\label{channel}
\end{equation}
is a promising channel to search for the charged Higgs boson at the Tevatron.
Simulations are performed at the centre-of-mass energy $\sqrt s=1960$~GeV and for $\tan\beta=20$. 

\clearpage
\section{COMPARISON OF PRODUCTION CROSS SECTIONS}
\vspace*{-0.1cm}

\begin{figure}[h]
\begin{minipage}{0.4\textwidth}
\includegraphics[width=1\textwidth]{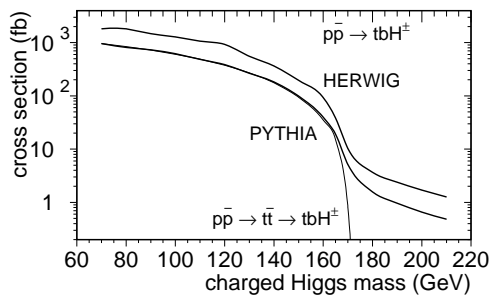}
\end{minipage} \hfill
\begin{minipage}{0.55\textwidth}
\vspace*{-1.4cm}
\caption{\label{fig:xsec}
         Charged Higgs boson production cross section at $\sqrt s=1960$~GeV
         for $\tan\beta=20$. 
         For the $\rm p\bar p\rightarrow tbH^\pm$ process (thick lines),
         the HERWIG expectation is larger by about a
         factor 2 compared to PYTHIA because of the different default setups
         as described in the text. 
The differences between the two PYTHIA curves for the 
$\rm p\bar p\rightarrow tbH^\pm$ and $\rm  p\bar p\rightarrow t\bar t \to  tbH^\pm$
processes instead result from 
top decay width effects and because of 
the additional diagrams which do not proceed via direct $\rm t\bar t$ production.}
\end{minipage}
\vspace*{-1.1cm}
\end{figure}

The expected production cross sections are determined using HERWIG and PYTHIA simulations, 
and are shown in Fig.~\ref{fig:xsec}. The default mass and coupling parameters of 
HERWIG version 6.5 and PYTHIA version 6.2
are used.
The cross sections depend strongly on the top decay width over the 
investigated mass range.
For the top width, the Standard Model (SM) value 
$\rm \Gamma_{\rm t} = 1.53$~GeV is used at $m_{\rm H^\pm}=210$~GeV and the width is increased 
as a function of the charged Higgs boson mass
to $\rm \Gamma_{\rm t} =1.74$~GeV at $m_{\rm H^\pm}=70$~GeV in both generators.
The production cross section in HERWIG is about a factor 2 larger compared to PYTHIA 
which can be attributed to the default choices of the standard parameters.
It is mostly driven by the different choice of the heavy quark masses entering the Higgs-quark 
Yukawa coupling.
In PYTHIA a running  b-mass is used at the tbH$^+$-vertex. For $m_{\rm H^\pm}=150$ GeV the
b-quark mass of 4.80~GeV is reduced to $m_{\rm b}=3.33$ GeV, 
while HERWIG uses $m_{\rm b}=4.95$~GeV both in the kinematics and at the vertex.
Other relevant parameters are the default Parton Distribution
Functions (PDFs) and the coupling constants $\alpha$ and $\alpha_s$,
as well as the scales used for evaluating the PDFs and couplings,
which are not the same in the default setups of the two simulation packages.

Tests comparing the total cross sections from
HERWIG and PYTHIA for {\em identical choices} of all
above parameters were performed and confirmed that the
two implementations of the hard scattering matrix elements coincide
numerically. In this study, however, we maintain the default configurations
of the two simulation packages. Hence, differences in the various 
distributions may be taken as an indication
of the theoretical systematic errors affecting this process. 

\section{COMPARISON OF BASIC SELECTION VARIABLES}
\vspace*{-0.1cm}

\begin{figure}[thp]
\vspace*{-0.1cm}
\begin{center}
\includegraphics[width=3.9cm]{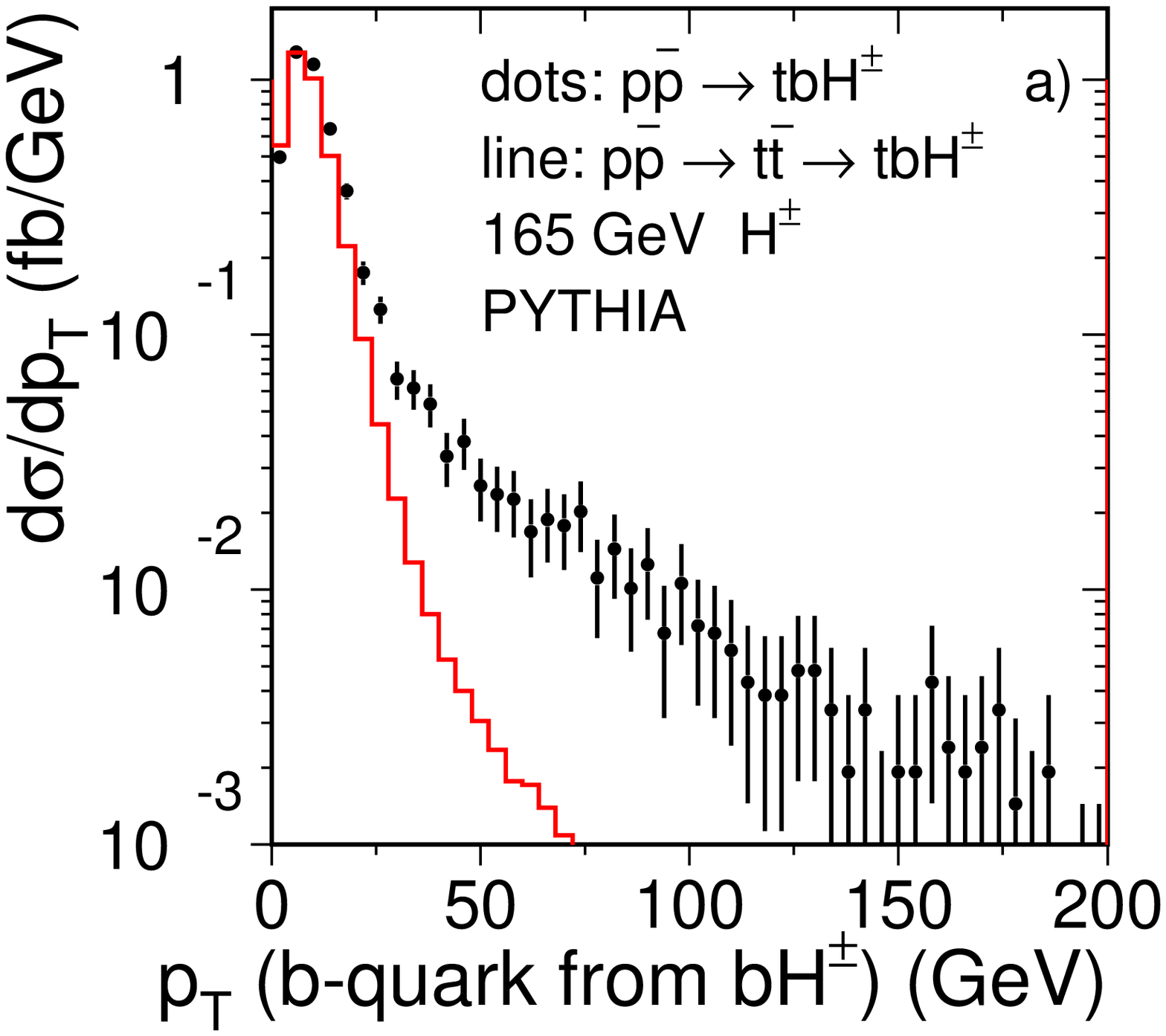}\hfill\includegraphics[width=3.9cm]{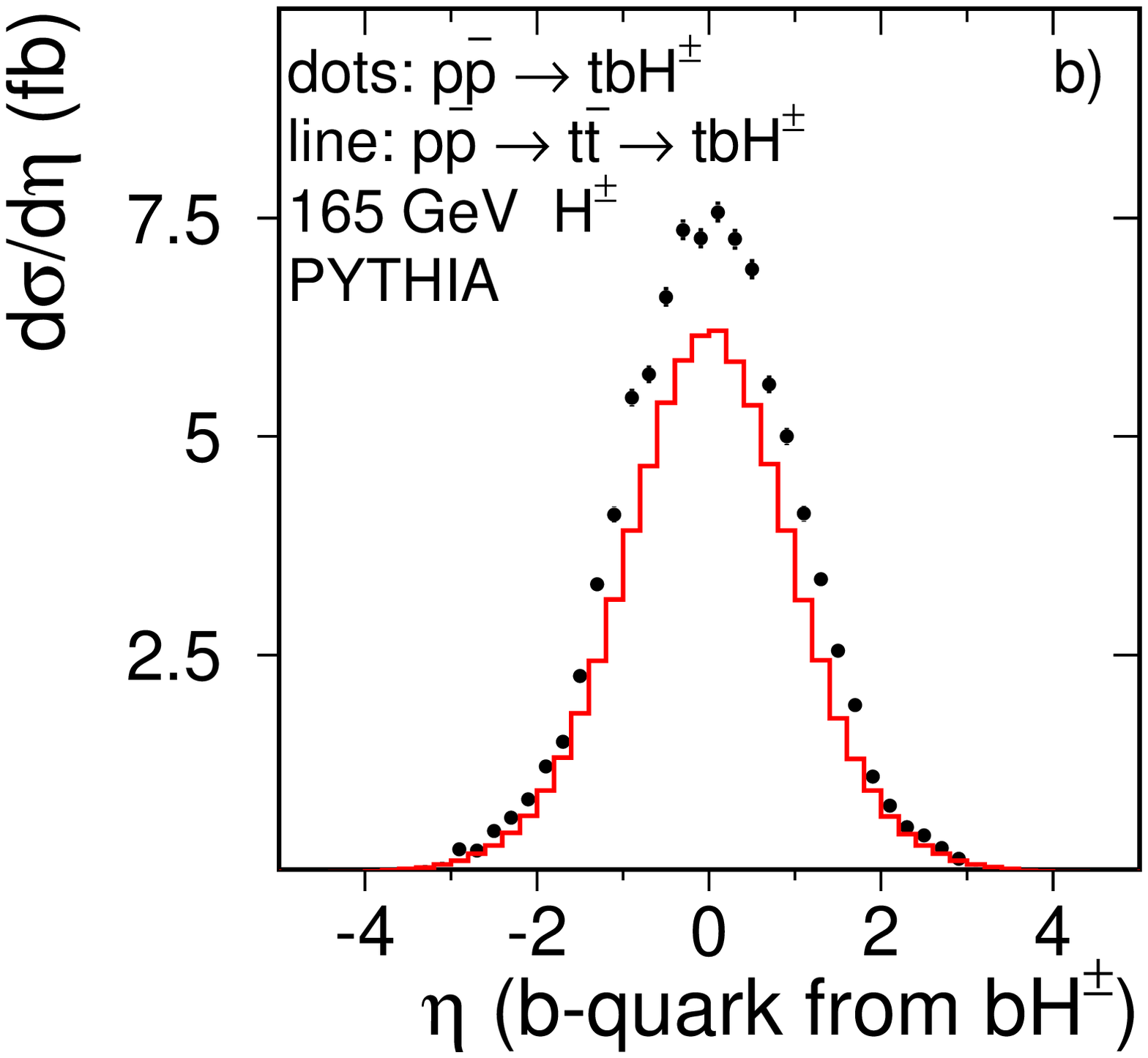}\hfill
\includegraphics[width=3.9cm]{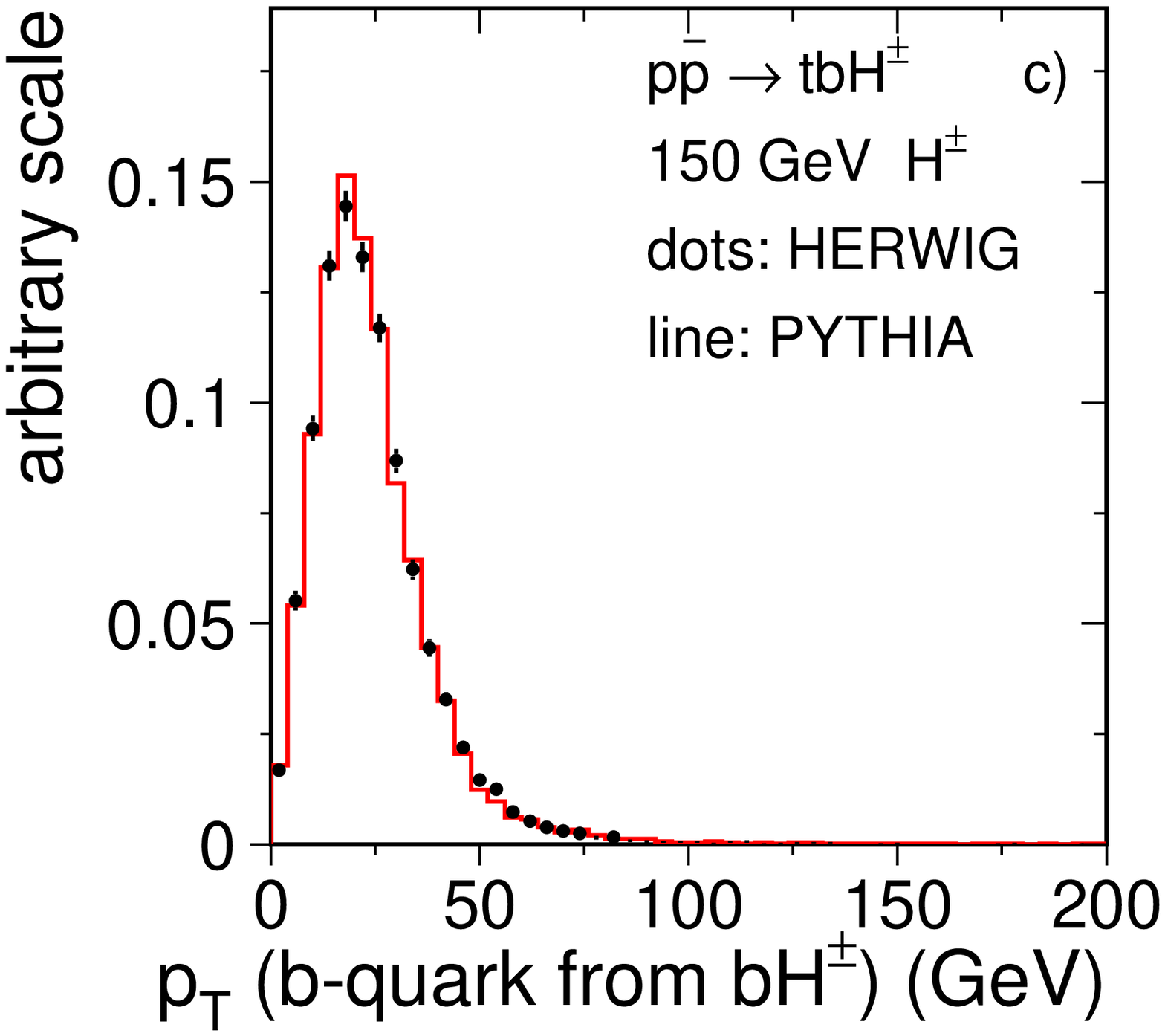}\hfill\includegraphics[width=3.9cm]{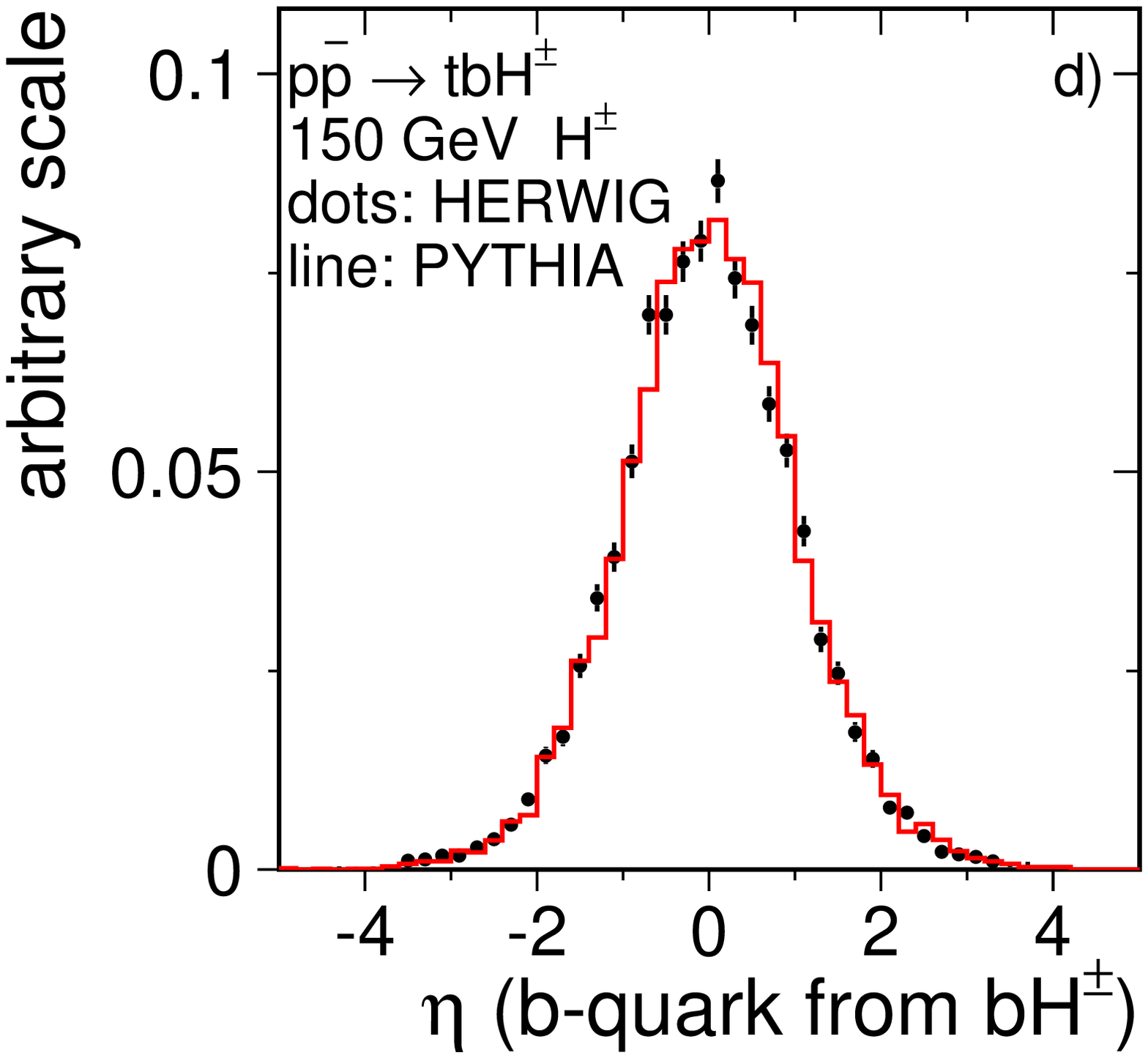}\\
\includegraphics[width=3.9cm]{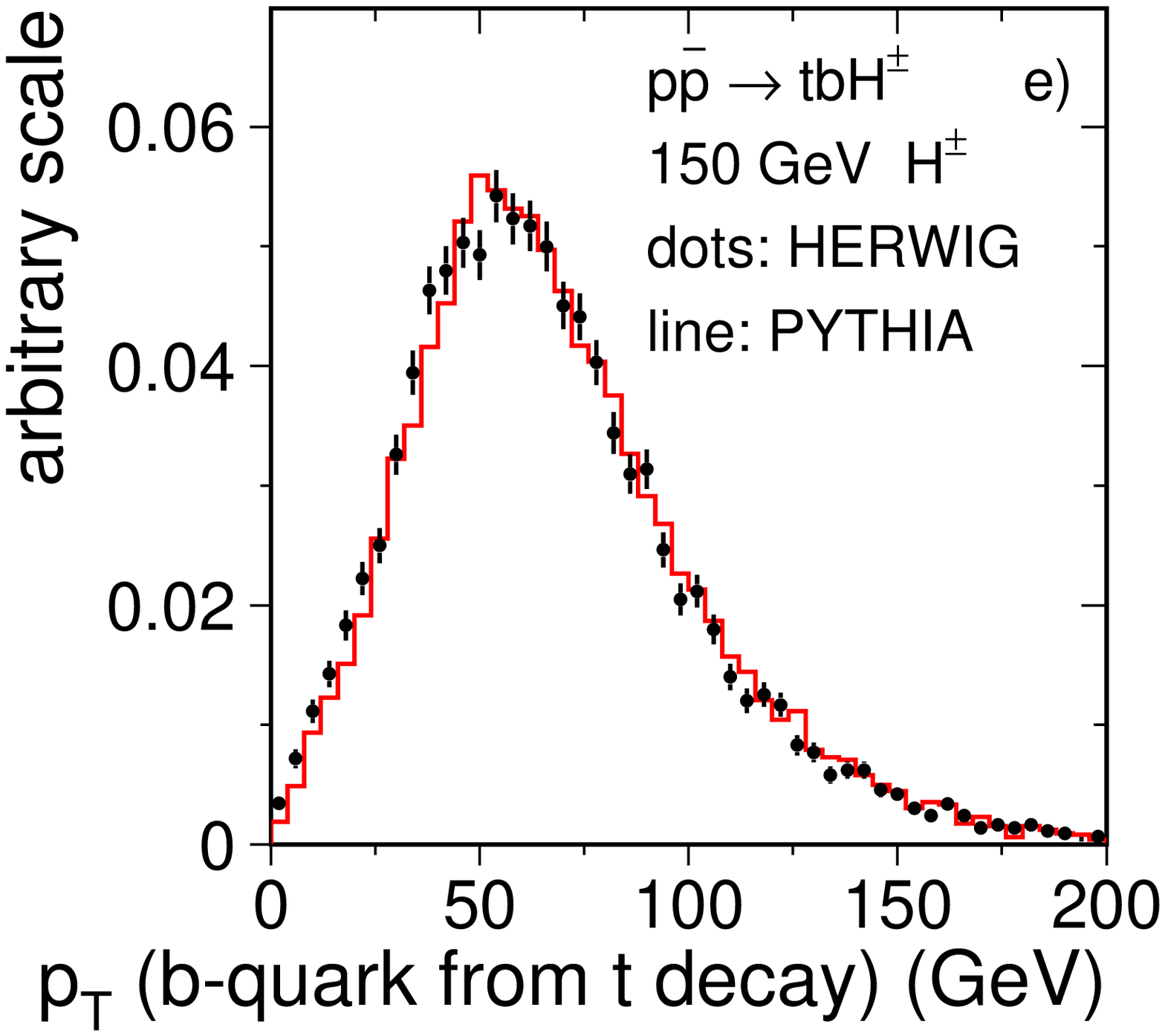}\hfill\includegraphics[width=3.9cm]{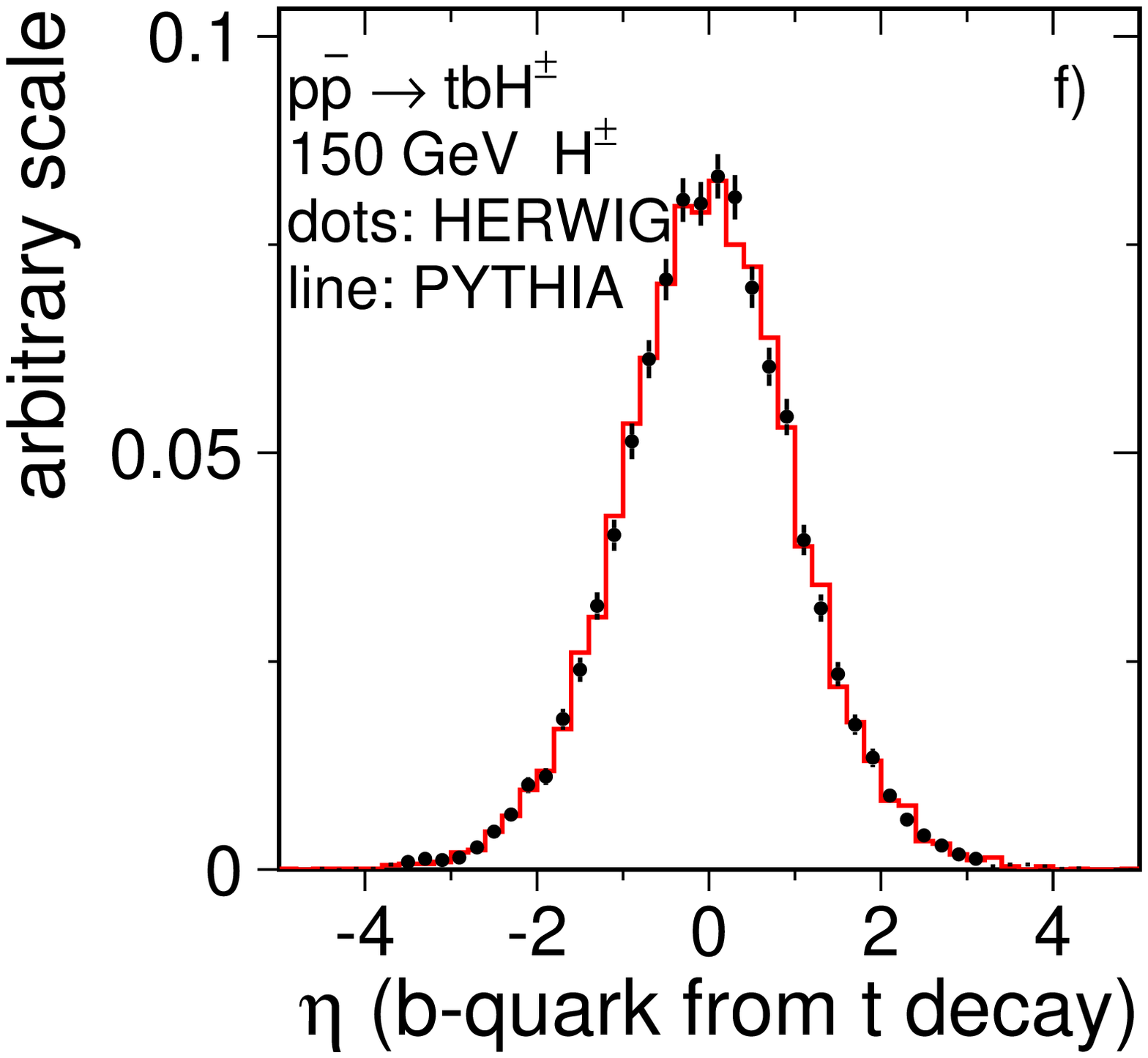}\hfill
\includegraphics[width=3.9cm]{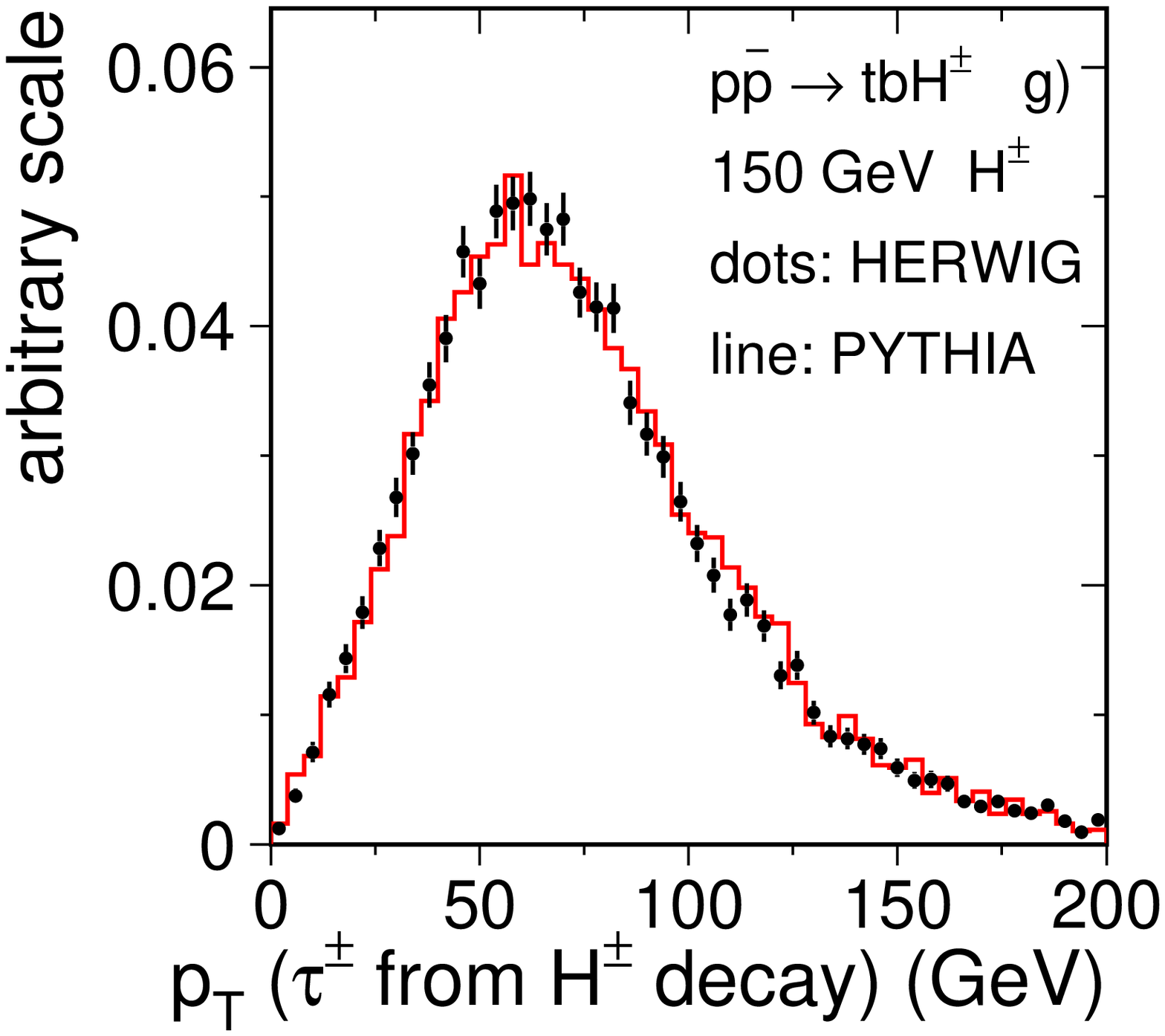}\hfill\includegraphics[width=3.9cm]{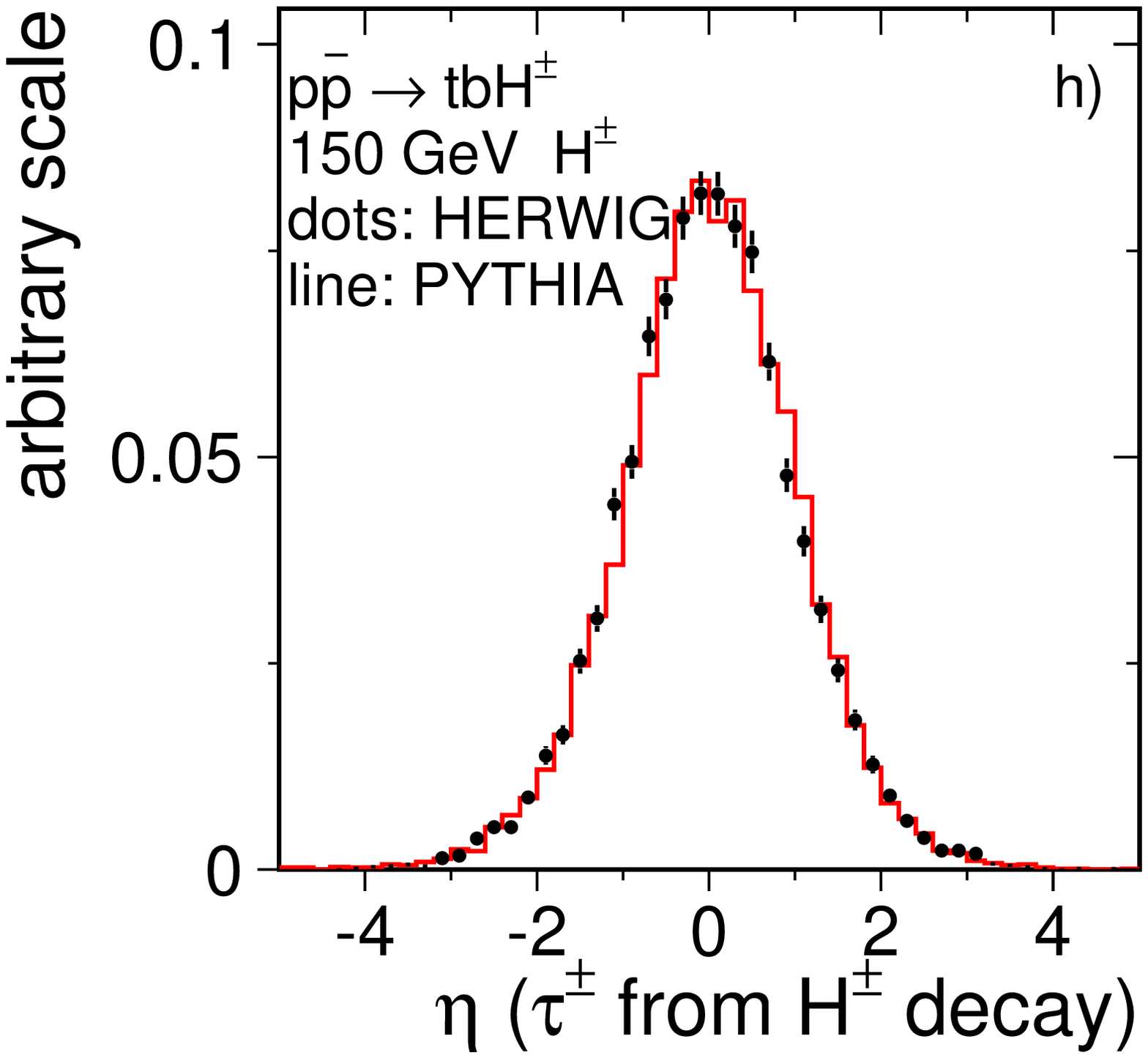}\\
\vspace*{-0.4cm}
\caption{\label{fig:distributions}
         Distributions of charged Higgs boson selection variables at the parton level
         for $\sqrt s=1960$~GeV and $\tan\beta=20$ for HERWIG and PYTHIA. 
         The variables are described in the text. 
%         In a) and b) the dominant effect is from the top off-shellness.
In a) and b) the differences are mainly from top off-shellness effects.
         In c) to h) each pair of curves is normalised to an equal area. 
         The error bars on the dots indicate the statistical\,uncertainty.
}
\end{center}
\end{figure}

At the parton level, several distributions of variables related to the 
event topology are compared between HERWIG and PYTHIA simulations.
In addition, differences in the distributions between the $\rm p\bar p\rightarrow tbH^\pm$ process
and the $\rm  p\bar p\rightarrow t\bar t \to  tbH^\pm$ subprocess are demonstrated.
Each comparison is based on two samples of 10,000 generated events.
Effects of the different event fragmentation schemes in HERWIG and PYTHIA could influence the 
comparison and they are not considered here.
The detector simulation of the Tevatron experiments would reduce further the
sensitivity of these comparisons. 
Figure~\ref{fig:distributions} shows the transverse momentum $p_{\rm T}$ and
pseudorapidity $\eta$ of the following particles:
\smallskip
\begin{itemize}
\item[a), b)] The b-quark produced in association with the $\rm H^\pm$ 
               in the $\rm p\bar p \rightarrow tbH^\pm$ (dots) and
               $\rm p\bar p \rightarrow t\bar t \rightarrow tbH^\pm$ (solid line) processes
               in PYTHIA  
               for $m_{\rm H^\pm} = 165$~GeV.
\item[c), d)] The b-quark produced in association with the $\rm H^\pm$
               in the $\rm p\bar p \rightarrow tbH^\pm$ process
               in HERWIG (dots) and PYTHIA (solid line)
               for $m_{\rm H^\pm} = 150$~GeV.
\item[e), f)] The b-quark from the top quark decay ($\rm t\rightarrow bW^\mp$) 
               in the $\rm p\bar p \rightarrow tbH^\pm$ process
               in HERWIG (dots) and PYTHIA (solid line)
               for $m_{\rm H^\pm} = 150$~GeV.
\item[g), h)] The $\tau$ lepton from the $\rm H^\pm$ decay 
               in the $\rm p\bar p \rightarrow tbH^\pm$ process 
               in HERWIG (dots) and PYTHIA (solid line)
               for $m_{\rm H^\pm} = 150$~GeV.
\end{itemize}
\smallskip
The differences in the $p_{\rm T}$ and $\eta$ distributions
are clearly visible in Figs.~\ref{fig:distributions}~a)~and~b) 
between the processes 
$\rm p\bar p \rightarrow tbH^\pm$ and 
$\rm p\bar p \rightarrow t\bar t \rightarrow tbH^\pm$.
The HERWIG and PYTHIA simulations show good agreement in the kinematic 
distributions of Figs.~\ref{fig:distributions}~c)~to~h) 
for both b-quarks and the $\tau$ lepton.
The decay of the $\tau$ lepton is not considered here, 
but it should be noted that
spin correlations must be taken into account in the study of the final state 
particles~\cite{Guchait:2001pi,spin95}.
\vspace*{-0.3cm}

\clearpage

\vspace*{-0.8cm}
\section{CONCLUSIONS}
At Tevatron Run-II,
about 1000 $\rm p\bar p \to tbH^\pm$
events per 1~fb$^{-1}$ at $\sqrt s=1960$~GeV  
could be produced for $m_{\rm H^\pm} = 100$~GeV and $\tan\beta=20$, while about 100 events are expected for 
$m_{\rm H^\pm}=150$~GeV.
These expected event rates will strongly be reduced when selection criteria
are applied to separate signal and background events.
For the default choices of mass and coupling parameters
in HERWIG and PYTHIA we observe significant differences
in the simulated total cross sections.
We have also studied the shape of basic selection variable distributions and found 
good agreement between
the HERWIG and PYTHIA parton level predictions in the $\rm p\bar p \to tbH^\pm$ process.
In comparison with the $\rm p\bar p \rightarrow t\bar t \to tbH^\pm$ subprocess, 
which was used in previous HERWIG and PYTHIA versions, for $m_{\rm H^\pm} > 160$~GeV the 
simulation of the full process results in significantly different distributions of $\rm tbH^\pm$ 
selection variables, mainly in the $p_{\rm T}$ distribution of the b-quark produced in association 
with the $\rm H^\pm$.

\vspace*{-0.1cm}
\section*{ACKNOWLEDGMENTS}

We would like to thank the Les Houches conference organizers
for their kind invitation and Nils Gollub for help with the 
tests comparing HERWIG and PYTHIA for identical choices of parameters.
SM thanks the Royal Society and AS the Particle Physics and Astronomy Research Council 
for financial support.

\vspace*{-0.1cm}
\bibliography{proc}
\vspace*{-0.8cm}
\mbox{ }
\end{document}